\documentclass[11pt]{article}

\usepackage{graphicx}                  
\usepackage{dcolumn}                   
\usepackage{bm}

\usepackage{jheppub}                   

\makeatletter
\gdef\@fpheader{\ }                    
\makeatother

\usepackage{graphicx}
\usepackage{dcolumn}
\usepackage{bm}
\usepackage{float}

\usepackage{amsmath,amssymb}           
\usepackage{amscd}                     
\usepackage{epsfig}                    
\usepackage[matrix,arrow]{xy}          
\usepackage{xspace}                    
\usepackage{stmaryrd}                  
\usepackage{slashed}
\usepackage{tikz}
\usepackage{hyperref}

\DeclareSymbolFont{bbold}{U}{bbold}{m}{n}
\DeclareSymbolFontAlphabet{\mathbbold}{bbold}

\newcommand{\bq}{\begin{equation}}
\newcommand{\eq}{\end{equation}}
\newcommand{\bea}{\begin{eqnarray}}
\newcommand{\eea}{\end{eqnarray}}

\newcommand{\dd}{\mathrm{d}}
\newcommand{\ee}{\mathrm{e}}
\newcommand{\ii}{\mathrm{i}}

\newcommand{\der}{\partial}

\newcommand{\bbR}{\mathbb{R}}

\DeclareMathOperator{\SU}{\mathit{SU}}
\DeclareMathOperator{\SO}{\mathit{SO}}

\DeclareMathOperator{\GL}{\mathit{GL}}

\DeclareMathOperator{\Symp}{\mathit{USp}}
\DeclareMathOperator{\Spin}{\mathit{Spin}}

\DeclareMathOperator{\Cliff}{Cliff}

\newcommand{\rep}[1]{\mathbf{#1}}
\newcommand{\repp}[2]{(\rep{#1}, \rep{#2})}

\newcommand{\LgenS}{L^{{}^{\text{KD}}}}

\newcommand{\Dgen}{{D}}

\newcommand{\LC}{\nabla}

\newcommand{\GenRic}{R^{\scriptscriptstyle 0}}

\newcommand{\GenS}{R}

\newcommand{\proj}[1]{\times_{#1}}

\DeclareMathOperator{\Edd}{\mathit{E_{d(d)}}}

\DeclareMathOperator{\dHd}{\mathit{\tilde{H}_d}}

\newcommand{\tF}{{\tilde{F}}}

\newcommand{\Tgen}{W}		 	
\newcommand{\Tint}{\Tgen_{\text{int}}}

\newcommand{\repsilon}{\hat{\epsilon}}

\newcommand{\tU}{{\tilde{U}}}
\newcommand{\tu}{{\tilde{u}}}

\newcommand{\hs}[1]{\hspace{#1}}

\newcommand{\ra}{\rightarrow}

\newcommand{\cN}{\mathcal{N}}
\newcommand{\AdSR}{G_R}

\title{
Supersymmetric AdS backgrounds and weak generalised holonomy
}

\author[a]{Andr\'e Coimbra,}
\emailAdd{andre.coimbra@ipht.fr}
\author[b]{Charles Strickland-Constable}
\emailAdd{charles.strickland-constable@ed.ac.uk}

\affiliation[a]{Institut de physique th\'eorique, 
	Universit\'e Paris Saclay, CEA, CNRS, F-91191 Gif-sur-Yvette, France}

\affiliation[b]{School of Mathematics and Maxwell Institute for Mathematical Sciences, University of Edinburgh, KingÕs Buildings, Edinburgh EH9 3FD, UK
}

\subheader{\textrm{EMPG-17-17 \\ IPhT-T17/152}
}

\abstract{
We study supersymmetric AdS$_D$ backgrounds of eleven-dimensional or type II supergravity preserving $\mathcal{N}$ supersymmetries using generalised geometry. 
We show that a large class correspond precisely to spaces admitting a generalised $G_{D,\mathcal{N}}$ structure with a weak integrability condition, which we call ``weak generalised holonomy".
This class contains all such backgrounds with odd $D$, but for even $D$ we must impose a technical assumption concerning the inner products of Killing spinors. 
It is known that all compact smooth solutions are included, though our analysis is purely local.  We show that the Killing superalgebras of such backgrounds are isomorphic to the standard $\mathcal{N}$-extended AdS$_D$ superalgebras, and give a geometric proof of the existence of isometries of the internal spaces which realise the AdS R-symmetry, as often assumed in explicit studies of such solutions.
\vfill}

\begin{document}  
 
\maketitle


\section{Introduction}
\label{sec:intro}

Generalised geometry~\cite{Hitchin:2004ut,Gualtieri:2003dx} has emerged as a powerful tool to address previously intractable problems in supergravity theories. In particular, it has enabled a much wider understanding of valid Anti de-Sitter backgrounds of string and M theory which preserve supersymmetry, which is crucial for the development of the AdS/CFT programme~\cite{Maldacena:1997re}. For instance, the rewriting of the differential conditions for preserving supersymmetry into the more elegant language of $O(d,d)$ generalised geometry~\cite{GMPT} was applied directly to specific AdS/CFT problems in~\cite{Minasian:2006hv,Gabella:2009wu,Gabella:2010cy}, and
was a key first step~\cite{Tomasiello:2011eb} for the works of~\cite{GCG-AdS7,GCG-AdS6,Tomasiello-other,Passias:2017yke} which include a complete classification of all possible AdS$_7$ solutions. The $\Edd\times\bbR^+$ version of generalised geometry which geometrises both NSNS and RR sectors~\cite{chris,PW,CSW2,CSW3} has been applied, for example, to the study of consistent truncations preserving maximal supersymmetry in~\cite{Lee:2014mla}, half-maximal in~\cite{Malek1,Malek2}, or in a slightly modified language for massive IIA in~\cite{CFPSW,Ciceri:2016dmd}, while work started in~\cite{GLSW,Ashmore:2015joa} made it possible to then begin characterising in full generality the  moduli space of $\mathcal{N}=2$ AdS backgrounds~\citep{Ashmore:2016qvs,Grana:2016dyl}, with direct applications to the understanding of the marginal deformations of the CFT duals in~\cite{Ashmore:2016oug}. This last set of results relied on a crucial integrability condition that supersymmetric AdS backgrounds must satisfy, formulated in generalised geometry, and which turns out to be a natural generalisation of the notion of Sazaki-Einstein or weak-$G_2$ spaces. It is this condition that will be the main focus of this paper.

The basic objects that we will study are $G$-structures in exceptional generalised geometry~\cite{PW,GO1,GO2}. 
$G$-structures and intrinsic torsion have long been employed to study supersymmetric backgrounds, since the pioneering work of~\cite{Gauntlett:2002sc}, and many of the early systematic studies of AdS flux backgrounds followed this approach~\cite{Lukas:2004ip,Behrndt:2004bh,Gauntlett:2004zh,Lust:2004ig,Gauntlett:2005ww}. However, while in ordinary differential geometry the inclusion of fluxes introduces intrinsic torsion, which breaks the integrability of the structure, there is an improved situation for $G$-structures in generalised geometry.

It was first shown in~\cite{CSW4} that generic flux backgrounds of Minkowski space with $\mathcal{N}=1$ supersymmetry must be \emph{generalised special holonomy} spaces, that is, they admit a reduced generalised structure group whose generalised intrinsic torsion~\cite{CSW4,CMTW} vanishes (see also~\cite{GO1,GO2}). This was further extended in~\cite{CS2} to show that if the background preserves more supersymmetry, then there exists 
a generalised torsion-free $G_{\mathcal{N}}$-structure where the group $G_{\mathcal{N}}$ is unique for each $\mathcal{N}$, see table~\ref{table1}.

\begin{table}[htb]
\centering
\begin{tabular}{lll}
   $d$ & $\dHd$ &  $G_{\mathcal{N}}$  \\ 
   \hline
   7 & $\SU(8)$ &  $\SU(8-\mathcal{N})$\\
   6 & $ \Symp(8)$ & $\Symp(8-2\mathcal{N})$ \\
   5 & $\Symp(4)\!\! \times \!\! \Symp(4)$ &   
      $\Symp(4-2\mathcal{N}_+)\!\times\!\Symp(4-2\mathcal{N}_-)$\\
   4 & $\Symp(4)$ & $\Symp(4-2\mathcal{N}) $
\end{tabular}
\caption{Local symmetry groups $\dHd$ of the generalised tangent space in $\Edd\times\bbR^+$ generalised geometry, and the
generalised special holonomy groups $G_{\mathcal{N}}\subset \dHd$ preserving $\mathcal{N}$ supersymmetry in $(11-d)$-dimensional Minkowski backgrounds. Note that for $d = 5$ we have six-dimensional supergravity with $(\mathcal{N}_+,\mathcal{N}_-)$ supersymmetry.}
\label{table1}
\end{table}

For the AdS case, it was shown in~\cite{CS1} that generic $\mathcal{N}=1$ backgrounds with flux correspond to a weakening of this integrability condition -- the torsion is not vanishing, instead it is set to a constant singlet which is the cosmological constant $\Lambda$. While the generalised special holonomy spaces of Minkwoski backgrounds are generalised Ricci flat, the supersymmetric AdS ones are generalised Einstein spaces
\begin{equation}
\GenRic = 0, \quad \GenS = (9-d)(10-d)\Lambda^2 ,
\end{equation}
where $\GenRic$ is the traceless part of the generalised Ricci tensor and $\GenS $ is the generalised Ricci scalar~\cite{CSW2}, and therefore automatically solve the equations of motion~\cite{Gauntlett:2002fz}. 

In this paper we will show that general $\mathcal{N}$-supersymmetric AdS backgrounds can also be described as $G_{\mathcal{N}}$ spaces with singlet torsion, with a subtle yet important caveat. We must assume that the Killing spinors which generate the supersymmetries may be taken to be orthonormal. This is a point that is often understated in the literature (for example it was overlooked in the generalised intrinsic torsion constructions of~\cite{Ashmore:2016qvs,Grana:2016dyl,Malek1,Malek2}), but unlike what happens in the Minkwoski case, the Killing spinor equations do not actually impose orthonormality for AdS$_D$ backgrounds with even $D$. In~\cite{Gabella:2012rc} it was shown that for AdS$_4$ it is sufficient to assume that solutions are complete and regular to ensure this property of the spinors, and so despite our assumption we can cover a large class of backgrounds which are interesting for applications. We hope to return to this issue in the near future, but for now we will just restrict our analysis to backgrounds with $\mathcal{N}$ orthonormal Killing spinors. We review all these points in section~\ref{sec:sugra}.

Having said that, we do indeed find that such AdS backgrounds are in precise correspondance to spaces with \emph{weak $G_{\mathcal{N}}$ special holonomy}, with a singlet torsion given by the constant $\Lambda$ which also fixes the AdS R-symmetry. This follows almost imediately from the results of~\cite{CS2} and~\cite{CS1}, as we show in section~\ref{sec:gen-hol}.

Finally, as in~\cite{CS2}, the proof of the torsion conditions for the $G_{\mathcal{N}}$ AdS structures requires evaluating the Kosmann-Dorfman derivative of the Killing spinors, the analogue of Kosmann's spinorial Lie derivative in generalised geometry. 
This allows us to derive an algebra generated by the Killing spinors on the internal space, which in turn fixes the form of the higher-dimensional Killing superalgebra first introduced in~\cite{FigueroaO'Farrill:2004mx}.
Furthermore, just like it was possible to prove that Minkwoski backgrounds must admit a number of commuting generalised isometries by writing this algebra in generalised geometry language, we will also find here that the AdS supersymmetry conditions imply the existence of generalised isometries that generate the R-symmetry groups. However, in contrast to the Minkowski case, for AdS it is possible to show that these generalised isometries necessarily give rise to genuine ordinary isometries of the backgrounds obeying the same algebra. This is something that is often assumed to be the case in AdS/CFT ansatze, but here we will show that it is in fact a necessity, for the class of backgrounds we are considering. This will be subject of section~\ref{sec:ksa}.\footnote{On completion of this work, we became aware of~\cite{Beck:2017wpm}, which
has substantial overlap with our results in section~\ref{sec:ksa}, via different methods.
}


\section{AdS backgrounds}
\label{sec:sugra}

We work in the exact same setup as in~\cite{CS1}. We decompose the higher dimensional spacetime under the warped metric
\begin{equation}
\label{eq:g}
   \dd s^2 = \ee^{2A}\dd s^2(\text{AdS}_D) + \dd s^2(M) , 
\end{equation}
and only consider the cases where the external AdS factor has dimension $D\geq 4$. The internal spin manifold $M$ is then of dimension $d$ in M theory and $d-1$ in type II, so we have that $d=11-D$. Our bosonic degrees of freedom are the internal Riemannian metric $g$, the warp factor $A$ which is a scalar function of the internal coordinates and the components of the fluxes which are consistent with the $D$-dimensional AdS symmetry. To match the conventions of~\cite{CSW2,CSW3}, we take $A=\Delta$ in M theory, so that $A=\Delta+\frac{1}{3}\phi$ in type II, where $\phi$ is the dilaton and the metric is in the string frame. Fermion fields are set to zero and we work in the supergravity limit $\alpha'=0$.

\subsection{Killing spinor equations}
\label{sec:KSE}

In a supersymmetric AdS space, after making a particular choice of how to break the Minkowski R-symmetry~\cite{CS1}, one has Killing spinors $\eta$ which must satisfy
\begin{equation}\label{eq:ads-kse}
\begin{aligned}
	\LC_\mu \eta_{\pm} &= \tfrac12 \Lambda \gamma_\mu  \eta_{\mp}, &\quad \text{in } D = 4,\\
	\LC_\mu \eta^A &= \tfrac12 (\sigma^3)^A{}_B \Lambda \gamma_\mu  \eta^B, &\quad \text{in } D = 5,\\
	\LC_\mu \eta_\pm^A &= \tfrac12  \Lambda \gamma_\mu  \eta_\mp^A, &\quad \text{in } D = 6,\\
	\LC_\mu \eta^A &= \tfrac12 \Lambda \gamma_\mu  \eta^A, &\quad \text{in } D = 7,\\
\end{aligned}
\end{equation}
where $\LC$ is the Levi--Civita connection in $\text{AdS}_D$, $A,B$ are the $\SU(2)$ indices of the symplectic Majorana and Majorana-Weyl spinors in $D=5,7$ and $D=6$ respectively, and $\pm$ subscripts denote chirality in even dimensions under the top gamma matrix $\gamma^{(D)}$.\footnote{For more details on conventions on Clifford algebras, intertwiners, spinor decompostions, etc. please refer to appendix \textbf{B} of~\cite{CSW3}. Here, for convenience, we choose a representation in which $(\gamma_m)^T = (\gamma_m)^* = - \gamma_m$ for $d=6,7$.} We must now tensor these with the internal Killing spinors to obtain a supersymmetric solution of the full higher-dimensional theory.  For concreteness we will focus on the M theory description, though our analysis covers the type II cases straightforwardly~\cite{CSW2,CSW3,CSW4}. 

Given spinors $\epsilon$ on the internal space $M$, we construct an eleven-dimensional spinor $\varepsilon^-$ as 
\begin{equation}
\begin{aligned}
	\varepsilon^- &= \eta_+ \otimes \epsilon + \eta_- \otimes \epsilon^*, &\quad \text{in } D = 4,\\
	\varepsilon^- &= \epsilon_{AB} \, \eta^A \otimes \epsilon^B, &\quad \text{in } D = 5,\\
	\varepsilon^- &= \epsilon_{AB} \, \eta_+^A \otimes \epsilon_+^B 
		+ \epsilon_{AB} \, \eta_-^A \otimes \epsilon_-^B, &\quad \text{in } D = 6,\\
	\varepsilon^- &= \epsilon_{AB} \, \eta^A \otimes \epsilon^B, &\quad \text{in } D = 7.\\
\end{aligned}
\end{equation}
Using the definitions of~\cite{CSW3} for the fermionic fields, we find the internal components of the Killing spinor equations for the eleven-dimensional supersymmetry parameter $\varepsilon^-$ can then be neatly written in all dimensions as
\begin{equation}
\label{eq:AdS-susy-ferm}
\begin{aligned}
   \Big[ 
      \LC_m + \tfrac{1}{288} F_{n_1 \dots n_4} \left(
         \Gamma_m{}^{n_1 \dots n_4} 
         - 8 \delta_{m}{}^{n_1} \Gamma^{n_2 n_3 n_4}\right) 
       - \tfrac{1}{12} \tfrac{1}{6!} \tilde{F}_{mn_1 \dots n_6} 
	 \Gamma^{n_1 \dots n_6} 
      \Big] \varepsilon^- & = 0,\\
        \Big[ 
      \slashed{\LC} - \tfrac{1}{4} \slashed{F} 
      - \tfrac{1}{4} \slashed{\tilde{F}} 
      + \tfrac{D-2}{2} (\slashed{\der} \Delta) 
      \Big] \varepsilon^- 
      + \tfrac{D-2}{2} \Lambda \ee^{-\Delta}\varepsilon^+ & = 0,
\end{aligned}
\end{equation}
where  $F, \,\tF$ are the internal four- and seven-form fluxes respectively, $\Gamma_{m}$ are now  $\Cliff(10,1)$ gamma matrices, and we define $\varepsilon^+$ (see also~\cite{Martelli:2003ki,Gauntlett:2004zh}) by
\begin{equation} \label{eq:epsilon-plus}
\begin{aligned}
	\varepsilon^+ &= \ee^{-2\ii\theta} \eta_+ \otimes \epsilon^* 
		+ \ee^{2\ii\theta} \eta_- \otimes \epsilon, &\quad \text{in } D = 4,\\
	\varepsilon^+ &= -\epsilon_{AC}(\sigma^3)^C{}_{B} \, \eta^A 
		\otimes ( \ii \gamma^{(6)} \epsilon^B), &\quad \text{in } D = 5,\\
	\varepsilon^+ &=  \delta_{AB}\, \eta_+^A \otimes \epsilon_-^B
		+ \delta_{AB} \, \eta_-^A \otimes \epsilon_+^B, &\quad \text{in } D = 6,\\
	\varepsilon^+ &= \epsilon_{AB} \, \eta^A \otimes (\gamma^{(4)}\epsilon^B), 
		&\quad \text{in } D = 7.\\
\end{aligned}
\end{equation}

We can now decompose the full eleven-dimensional Killing spinor equation to obtain the conditions on the internal spinor, and thus on the geometry of the internal manifold. 
For convenience we write only  the equations for the complex internal spinor $\epsilon$ and do not write the conjugate equations for $\bar\epsilon$ in the $D=4$ case; only the equations for the first ``half" $\epsilon \equiv \epsilon^1$ of the symplectic Majorana spinor $\epsilon^A$ (the ones for $\epsilon^2$ follow by conjugating those for $\epsilon^1$) in $D=5$; and also only those for $\epsilon \equiv \epsilon^1$ in the $D=7$ case.  As the AdS$_6$ Killing spinor equation can only be solved by a non-chiral six-dimensional spinor, we must have both $\eta_+^A$ and $\eta_-^A$ non-zero, and so we will be writing internal equations for (the first ``half" of) two symplectic Majorana spinors $\epsilon_+ = \epsilon_+^{A=1}$ and $\epsilon_- = \epsilon_-^{A=1}$.

Decomposing~\eqref{eq:AdS-susy-ferm} thus leads to the internal equations
\begin{equation}
\label{eq:ads-kse-int}
\begin{aligned}
\LC_m \epsilon + \tfrac{1}{288} 
   (\gamma_m{}^{n_1 \dots n_4} - 8 \delta_{m}{}^{n_1} \gamma^{n_2 n_3 n_4}) 
   F_{n_1 \dots n_4} \epsilon  - \tfrac{1}{12} \tfrac{1}{6!} \tilde{F}_{mn_1 \dots n_6} 			\gamma^{n_1 \dots n_6} \epsilon &= 0, \\
\slashed{\LC} \epsilon + \tfrac{9-d}{2} (\slashed{\der} \Delta) \epsilon - \tfrac{1}{4} \slashed{F}\epsilon - \tfrac{1}{4} \slashed{\tilde{F}} \epsilon  + \tfrac{9-d}{2} \ee^{-\Delta} X 
&= 0,
\end{aligned}
\end{equation}
where we have
\begin{equation}
\begin{aligned}
	X &= \Lambda \epsilon^* && \quad \text{in} \quad d=7 ,\\
	X &= - \ii \Lambda \gamma^{(6)} \epsilon && \quad \text{in} \quad d=6 ,\\
	X &= \Lambda \gamma^{(4)} \epsilon && \quad \text{in} \quad d=4 ,\\
\end{aligned}
\end{equation}
and for $d=5$ 
\begin{equation}
\label{eq:ads6-kse-int}
\begin{aligned}
\LC_m \epsilon_+ + \tfrac{1}{288} 
   (\gamma_m{}^{n_1 \dots n_4} - 8 \delta_{m}{}^{n_1} \gamma^{n_2 n_3 n_4}) 
   F_{n_1 \dots n_4} \epsilon_+  &= 0, \\
\LC_m \epsilon_- - \tfrac{1}{288} 
   (\gamma_m{}^{n_1 \dots n_4} - 8 \delta_{m}{}^{n_1} \gamma^{n_2 n_3 n_4}) 
   F_{n_1 \dots n_4} \epsilon_-  &= 0, \\
\slashed{\LC} \epsilon_+ + 2 (\slashed{\der} \Delta) \epsilon_+
	- \tfrac{1}{4} \slashed{F}\epsilon_+ 
	+ 2 \ee^{-\Delta} \Lambda \epsilon_-
&= 0 , \\
\slashed{\LC} \epsilon_- + 2 (\slashed{\der} \Delta) \epsilon_- 
	+ \tfrac{1}{4} \slashed{F}\epsilon_-
	- 2 \ee^{-\Delta} \Lambda \epsilon_+
&= 0 .
\end{aligned}
\end{equation}
If the background preserves $\mathcal{N}$ supersymmetries, one has $\mathcal{N}$ independent spinors $\epsilon_i$, for $i=1,\dots ,\mathcal{N}$, which satisfy these equations. In the AdS$_6$ case we can have either one or two pairs $(\epsilon_+)_i,(\epsilon_-)_i$ corresponding to $\mathcal{N}=(1,1)$ or $\mathcal{N}=(2,2)$ supersymmetry respectively. However, from the classification of maximally supersymmetric solutions~\cite{Figueroa-OFarrill:2002ecq}, there is only one $\mathcal{N}=(2,2)$ AdS$_6$ solution, which is simply a rewriting of AdS$_7\times S^4$ as a warped AdS$_6$ 
(see, for example,~\cite{Gran:2016zxk,Gutowski:2017edr}). Therefore, for the most part, we will only consider $\cN=(1,1)$ AdS$_6$ solutions in what follows.
In fact, due to many no-go, uniqueness and classification results the local supersymmetric AdS$_6$ and AdS$_7$ solutions are all explicitly known~\cite{GCG-AdS7,GCG-AdS6,Passias:2012vp,DHoker}.


\subsection{Inner products of Killing spinors}

Note that we can write the equations~\eqref{eq:ads-kse-int} in the equivalent form
\begin{equation}
\label{eq:ads-kse-int-equiv}
\begin{aligned}
	\LC_m \hat\epsilon  
	- \tfrac12 (\der_n \Delta) \gamma_m{}^n \hat\epsilon
	- \tfrac{1}{4} \tfrac{1}{3!} F_{mn_1 \dots n_3} \gamma^{n_1 \dots n_3} \hat \epsilon  
   	- \tfrac{1}{4} \tfrac{1}{6!} \tilde{F}_{mn_1 \dots n_6} \gamma^{n_1 \dots n_6} \hat\epsilon 
		&= \tfrac12\gamma_m \ee^{-\Delta}\hat{X}, \\
	(\slashed{\der} \Delta) \hat\epsilon 
		+ \tfrac{1}{6} \slashed{F}\hat\epsilon 
		+ \tfrac{1}{3}\slashed{\tilde{F}} \hat\epsilon  
		+  \ee^{-\Delta}\hat{X} &= 0,
\end{aligned}
\end{equation}
and the $d=5$ equations~\eqref{eq:ads6-kse-int}
\begin{equation}
\label{eq:ads6-kse-int-equiv}
\begin{aligned}
	\LC_m \hat\epsilon_+  
	- \tfrac12 (\der_n \Delta) \gamma_m{}^n \hat\epsilon_+
	- \tfrac{1}{4} \tfrac{1}{3!} F_{mn_1 \dots n_3} \gamma^{n_1 \dots n_3} \hat\epsilon_+  
		&= +\tfrac12\ee^{-\Delta} \Lambda \gamma_m \hat\epsilon_-, \\
	\LC_m \hat\epsilon_-
	- \tfrac12 (\der_n \Delta) \gamma_m{}^n \hat\epsilon_-
	+ \tfrac{1}{4} \tfrac{1}{3!} F_{mn_1 \dots n_3} \gamma^{n_1 \dots n_3} \hat\epsilon_-
		&= -\tfrac12\ee^{-\Delta} \Lambda \gamma_m \hat\epsilon_+, \\
	(\slashed{\der} \Delta) \hat\epsilon_+ 
		+ \tfrac{1}{6} \slashed{F}\hat\epsilon_+
		+  \ee^{-\Delta} \Lambda \hat\epsilon_- &= 0, \\
	(\slashed{\der} \Delta) \hat\epsilon_-
		- \tfrac{1}{6} \slashed{F}\hat\epsilon_-
		-  \ee^{-\Delta} \Lambda \hat\epsilon_+ &= 0 ,
\end{aligned}
\end{equation}
where we have rescaled
\begin{equation}
	\hat{\epsilon} = \ee^{-\Delta/2} \epsilon,
	\hs{40pt}
	\hat{X} = \ee^{-\Delta/2} X.
\end{equation}

We thus have that the $d=7$ Killing spinor equations define an $\SU(8)$ connection~\cite{CS2} (see also~\cite{DuffHull})
\begin{equation}
\label{eq:kse-su8-ord}
\begin{aligned}
	\tilde{\nabla}_m\hat\epsilon  = \LC_m \hat\epsilon  
	- \tfrac12 (\der_n \Delta) \gamma_m{}^n \hat\epsilon
	- \tfrac{1}{4} \tfrac{1}{3!} F_{mn_1 \dots n_3} \gamma^{n_1 \dots n_3} \hat\epsilon  
   	- \tfrac{1}{4} \tfrac{1}{6!} \tilde{F}_{mn_1 \dots n_6} \gamma^{n_1 \dots n_6} \hat\epsilon ,
\end{aligned}
\end{equation}
or more generically we have a $\dHd$ connection in $d$ dimensions (see table~\ref{table1}).  Recall~\cite{CSW3} that the group $\dHd$ can be defined from $\Cliff(d;\bbR)$ by its reversal involution $r:\gamma^{m_1\dots m_p}\mapsto \gamma^{m_p\dots m_1}$. One has that 
\begin{equation}
   \tilde{H}_{p,q} = \{ g\in\Cliff(p,q;\bbR) : r(g)g=1 \}
\end{equation}
For $p+q\leq 8$, the corresponding Lie algebras are then generated by $\{\gamma^{mn},
\gamma^{mnp},\gamma^{m_1\dots m_6},\gamma^{m_1\dots m_7}\}$.\footnote{Note that $\tilde{H}_{7,0}$
   is strictly $U(8)$. Dropping the $\gamma^{(7)}$ generator one gets
   $\tilde{H}_7=\SU(8)$.}

This makes it easy to evaluate the derivative of the inner product of any two spinors $\repsilon_i^\dagger \repsilon_j $ which satisfy the AdS-background Killing spinor equations. Using $\partial_m (\repsilon_i^\dagger \repsilon_j ) = \tilde{\nabla}_m (\repsilon_i^\dagger \repsilon_j )$ we find that for $d=4,6$ their inner product is necessarily constant. However, in  $d=7$ we have that the imaginary component of the derivative is non-vanishing
\begin{equation}
	\tilde{\nabla}_m (\repsilon_i^\dagger \repsilon_j ) = \tfrac12\Lambda\left( \repsilon_i^T \gamma_m \repsilon_j -( \repsilon_j^T \gamma_m \repsilon_i)^*\right).
\end{equation}
so, unlike for the Minkowski analysis, we cannot assume we have an orthonormal basis of Killing spinors. In fact, it is easy to check that in $d=7$ only the real combination
\begin{equation*}
	\repsilon_i^\dagger \repsilon_j + \repsilon_j^\dagger \repsilon_i = \text{constant},
\end{equation*}
and in AdS$_6$, $d=5$
\begin{equation}
	\repsilon_{+,i}^\dagger \repsilon_{+,j} + \repsilon_{-,i}^\dagger \repsilon_{-,j} = \text{constant}.
\end{equation}

As a side note, observe that it is the $\tilde{H}_{d+1}$ spinor norm that is preserved by the AdS background Killing spinor equations, which coincides with the $\dHd$ norm only for even $d$. That the $\tilde{H}_{d+1}$ inner product is preserved should come as no surprise, sinces by a cone construction AdS$_D$ backgrounds can always be thought of as warped Mink$_{D-1}$ backgrounds, whose internal geometry is in turn described by $\tilde{H}_{d+1}$ generalised geometry with orthornormal Killing spinors.

Now, as real linear combinations of Killing spinors give Killing spinors, the Killing spinors always form a real vector space. Therefore, by an orthogonal transformation and some rescalings, w.l.o.g. we can diagonalise the real component to
\begin{equation}
	\repsilon_i^\dagger \repsilon_j + \repsilon_j^\dagger \repsilon_i = 2\delta_{ij},
\end{equation}
and therefore take
\begin{equation}
\label{eq:ks-innerprod}
	\repsilon_i^\dagger \repsilon_j = \delta_{ij} + \ii S_{ij},
\end{equation}
where the $S_{ij} = S_{[ij]}$ are real functions. Note that for $\mathcal{N}=1$ the function $S_{ij}$ vanishes identically, and the Killing spinor norm $\repsilon^\dagger \repsilon$ is indeed always constant, as mentioned in~\cite{CS1}. 

The fact that $S_{ij}\neq 0$ in general, and therefore that the Killing spinors of an AdS background cannot be taken to be orthonormal, appears to be perhaps underappreciated in the literature. Of particular concern for us, the proof~\cite{CS2} that the Killing spinor equations constrain precisely the generalised intrinsic torsion leans heavily on the existance of an orthnormal basis of Killing spinors\footnote{It becomes clear that this point is important when one considers that without the orthonormality condition, there is no generalised $\dHd$ connection which preserves the Killing spinors.} (which is guaranteed for Minkwoski backgrounds). 
In~\cite{Gabella:2012rc}, the $\mathcal{N}=2$ (for which $S_{ij}$ has a single component), $d=4$ case was analysed in detail and it was concluded that complete regular solutions actually require $S_{ij} = 0$. However, more general solutions with $S_{ij}\neq0$ do exist. 
For example, any supersymmetric AdS$_5$ solution gives rise to a warped AdS$_4$ solution with  non-trivial functions $S_{ij}$, using the foliation of AdS$_5$ by AdS$_4$ leaves (see, for example,~\cite{Gran:2016zxk,Passias:2017yke} for recent accounts of this foliation).

In the following we will assume $S_{ij} = 0$ and leave for an upcoming paper~\cite{CS4} the discussion of the alternative.

\section{Generalised geometry analysis}
\label{sec:gen-hol}

\begin{table}[htb]
\begin{center}
\begin{tabular}{lccccc}
   $d$ & $\dHd$ & $\hat{S}^-$ & $\hat{S}^+$ & $\hat{J}^-$ & $\hat{J}^+$ \\
   \hline
   $7$ & $\SU(8)$ & $\rep{8}$ 
      & $\rep{\bar{8}}$ 
      & $\rep{56}$ 
      & $\rep{\bar{56}}$ \\
   $6$ & $\Symp(8)$ & $\rep{8}$ & $\rep{8}$ & $\rep{48}$ & $\rep{48}$ \\
   $5$ & $\Symp(4)\times\Symp(4)$ & $(\rep{4},\rep{1}) $ 
      & $(\rep{1},\rep{4})$ 
      & $(\rep{4},\rep{5}) $ 
      & $(\rep{5},\rep{4}) $ \\
   $4$ & $\Symp(4)$ & $\rep{4}$ & $\rep{4}$ 
      & $\rep{16}$ & $\rep{16}$ 
\end{tabular}
\end{center}
\caption{Double covers of the maximal compact subgroups of $\Edd\times\bbR^+$ and the representations of the bundles $S$ and $J$ corresponding to spinors and vector-spinors respectively in $\Spin(d)$. Note that $\Symp(2n)$ denotes the compact symplectic group of rank $n$.}
\label{table2}
\end{table}

Let us rewrite the AdS background KSEs in the compact language of generalised geometry, which makes their larger local $\dHd$ symmetry manifest. On the $d$-dimensional (or $(d-1)$-dimensional for type II) manifold $M$ we introduce the $\Edd\times\bbR^+$ generalised tangent bundle~\cite{PW,chris,CSW2}. There exists a natural differential strucutre given by the generalisation of the Lie derivative, the Dorfman derivative  $L_V$ along a generalised vector $V$, which generates infinitesimal diffeomorphisms and gauge transformations.

The background fields then determine a reduction of the structure group of the generalised tangent bundle to $\dHd$, the (double cover of the) symmetry group of the generalised metric $G$, which combines all bosonic fields in a single object. If one then considers generalised Levi--Civita~\cite{CSW2} connections $D$, that is connections with vanishing generalised torsion (defined in terms of the Dorfman derivative) and compatible with $\dHd$ such that $D G = 0$, one has that the Killing spinor equations take the form~\cite{CSW3}
\begin{equation}\label{eq:gen-kse}
\begin{aligned}
	\Dgen \proj{J^-} \hat\epsilon^- &= 0 ,\\
	\Dgen \proj{S^+} \hat\epsilon^- &= -\tfrac{9-d}{2} \Lambda \hat\epsilon^+,
\end{aligned}
\end{equation}
where $\proj{X}$ denotes projection to the $X$ representation and $S$ and $J$ are the representations of the spinor and the vector-spinor in $d$-dimensions respectively, see table~\ref{table2}. Note that, as mentioned in~\cite{CS1}, there are generically two ways of realising $\dHd$ in $\Cliff(d;\bbR)$, related by taking $\gamma^a\rightarrow -\gamma^a$, leading to two generically inequivalent spinor bundles $S^+$ and $S^-$ in odd dimensions (and therefore also two inequivalent $J^{\pm}$), while in even dimensions $-\gamma^{m}= \gamma^{(d)}\gamma^m(\gamma^{(d)})^{-1}$ and so $S^-\simeq S^+ $. For a discussion, see appendix \textbf{E} of~\cite{CSW3}.

While above we have written these equations in a formal generic notation, they can be given a precise form in explicit $\dHd$ indices, see~\cite{CS1}, which we will use below. 
We also stress that the formalism of generalised geometry describes equally well the type II supergravity theories. These simply have a different form of the generalised tangent space corresponding to the decompositions under the two inequivalent $\GL(d-1,\bbR)$ subgroups of $\Edd\times\bbR^+$. In particular, the equations of all three theories take precisely the same form when written in generalised geometry language. For example,~\eqref{eq:gen-kse} gives the Killing spinor equations in all cases. As such, all of our $\dHd$ covariant manipulations below apply equally well to the type II theories, and we can draw the same overall conclusions for those cases.


\subsection{$\SU(8)$ Killing spinors and generalised Killing vectors}

We now look at the implications of having $\mathcal{N}$ copies of the Killing spinor equations~\eqref{eq:gen-kse} from the point of view of generalised $G$-structures. As usual, we will focus on the M theory $d=7$ case. We thus have 
$\mathcal{N}$ independent non-vanishing spinors $\hat{\epsilon}_i^{\alpha}$ which are in the $\rep{8}$ of $\SU(8)$. Therefore, they define a reduction of the structure group of the generalised tangent bundle to $SU(8-\mathcal{N})\subset \SU(8)$~\cite{CSW4,CS2}. 

Furthermore they satisfy the Killing spinor equations in explict $\SU(8)$ indices~\cite{CS1}
\begin{subequations}
\label{eq:kse-su8}
\begin{align}
	(\Dgen \proj{J^-} \hat\epsilon_i^-){}^{[\alpha\beta\gamma]} &= \Dgen^{[\alpha\beta}\hat\epsilon_i^{\gamma]} = 0, \label{eq:kse-su8-grav} \\
	(\Dgen \proj{{S}^+} \hat\epsilon_i^-){}_\alpha 
		&= -\Dgen_{\alpha\beta} \hat\epsilon_i^\beta = - \Lambda (\bar{\hat\epsilon}_i)_\alpha .
		\label{eq:kse-su8-dil}
\end{align}
\end{subequations}

Now, we can use our set of spinors to build generalised vectors\footnote{Recall that in $d=7$, generalised vectors transform in the $\rep{56}_{\rep{1}}$ of $E_{7(7)}\times\bbR^+$, which under $\SU(8)$ decomposes to $\rep{28}+\bar{\rep{28}}$.} $V_{ij}$ and $W^{ij}$ as
\begin{equation}
\label{eq:gen-bilinears}
\begin{aligned}
	(V_{ij})^{\alpha \beta} &= \hat\epsilon^{[\alpha}_i \hat\epsilon^{\beta]}_j ,
	 & \hs{40pt} 
	(V_{ij})_{\alpha \beta} &= 0 ,\\
	(W^{ij})^{\alpha \beta} &= 0 ,
	 & \hs{40pt} 
	(W^{ij})_{\alpha \beta} &= \bar{\hat\epsilon}^i_{[\alpha} \bar{\hat\epsilon}^j_{\beta]} .
\end{aligned}
\end{equation}
Notice that generalised vectors are elements of $E\simeq TM\oplus\Lambda^2T^*M\oplus\dots$, so we see that from an ordinary geometry point of view these are composed of the usual $p$-form Killing spinor bilinears that feature in discussions of flux compactifications. For instance, we have that the (complex) vector component of $V_{ij}$ in the non-conformal split frame~\cite{CSW2,CS2} is 
\begin{equation}
	v^m_{ij} =  \tfrac{1}{32\ii} \epsilon_i^T \gamma^m \gamma^{(7)} \epsilon_j
		= - \tfrac{1}{32} \epsilon_i^T \gamma^m \epsilon_j ,
\end{equation}
where $\epsilon$ is the supergravity spinor without rescaling.

The Killing spinor equations~\eqref{eq:kse-su8} then imply that these generalised vectors satisfy differential conditions
\begin{equation}\label{eq:ks-gckv}
	D^{[\alpha\beta} V_{ij}^{\gamma\delta]} = 0,
	\hs{40pt}
	\bar{D}_{\alpha\beta} V_{ij}^{\alpha\beta} = -2\ii \Lambda S_{ij},
\end{equation}
where $S_{ij}$ was defined in~\eqref{eq:ks-innerprod}, and there are similar equations for $W_{ij}$. Thus for $S_{ij}=0$ we have that $V_{ij}$ and $W_{ij}$ are \emph{generalised Killing vectors}~\cite{CS2}, i.e. they satisfy the precise generalised analogue of $\nabla_{(m}v_{n)}=0$, which implies that $L_{V_{ij}}G=L_{W_{ij}}G=0$. If $S_{ij}\neq 0$ then it can be shown that they are instead generalised conformal Killing vectors~\cite{CS4}, but as mentioned we will not consider this case here, and we assume $S_{ij} = 0$.

Now, in~\cite{CS2} the generalised geometry version of Kosmann's spinorial Lie derivative was introduced, the Kosmann-Dorfman bracket $\LgenS$. For the $\SU(8)$ case at hand, it reads
for an arbitrary generalised vector $V=(V^{\alpha\alpha'}, \bar{V}_{\alpha\alpha'})$ and spinor $\zeta^{\alpha}$,
\begin{equation}
\begin{aligned}
\label{eq:gen-KD-su8}
	\LgenS_V \zeta^\alpha 
		= \tfrac{1}{32} &(V^{\gamma \gamma'} \bar\Dgen_{\gamma\gamma'} 
			+ \bar{V}_{\gamma \gamma'} \Dgen^{\gamma\gamma'}) \zeta^\alpha \\
			&- \tfrac{1}{16}\Big[ \Big( \bar\Dgen_{\beta \gamma} V^{\alpha \gamma}
				-  \Dgen^{\alpha \gamma} \bar{V}_{\beta \gamma} \Big)
			-\tfrac18 \delta^\alpha{}_\beta 
				\Big(\bar\Dgen_{\gamma \gamma'} V^{\gamma \gamma'}
				-  \Dgen^{\gamma \gamma'} \bar{V}_{\gamma \gamma'} \Big) \Big]
			\zeta^\beta .
\end{aligned}
\end{equation}

Crucially, if the generalised vector $V$ is generalised Killing, then
\begin{equation}
\label{eq:gen-comm-susy-lie}
	\LgenS_{V} \left( \Dgen \proj{S\oplus J}\zeta\right) = \Dgen \proj{S\oplus J} \left(\LgenS_{V} \zeta\right),
\end{equation}
and therefore the Kosmann-Dorfman derivative of a Killing spinor along a Killing vector is itself a Killing spinor. Since we have just seen that the AdS background Killing spinors define generalised Killing vectors~\eqref{eq:gen-bilinears} when $S_{ij}=0$, we have that
\begin{equation}\label{eq:KSA-coef}
	\LgenS_{V_{ij}} \hat\epsilon_k = X_{ijk}{}^l \hat\epsilon_l , 
\end{equation}
for some constant coefficients $X_{ijk}{}^l$, and similarly for $W^{ij}$. 


\subsection{Weak generalised special holonomy}
\label{sec:weak-gen-hol}

In~\cite{CS2} it was shown that there is an isomorphism between the space $\Tint$ of the generalised intrinsic torsion\footnote{Recall that the generalised intrinsic torsion of a generalised $G$-structure measures the obstruction to finding a generalised torsion-free connection that preserves that structure~\cite{CMTW,CSW4}.} of the $\SU(8-\mathcal{N})$ structure defined by the set of Killing spinors $\{\hat\epsilon_i\}$, and the space constrained by the Killing spinor equations plus an extra component that only exists for $\mathcal{N}> 2$
\begin{equation}
   \Tint \simeq \mathcal{N}\times(S \oplus J) \oplus \tbinom{\mathcal{N}}{3} \times V\oplus\text{c.c.} 
\end{equation}
Here $V$ is the bundle associated to the $\rep{[8-\mathcal{N}]}$ representation of $\SU(8-\mathcal{N})$.\footnote{The explicit representation that appear in $\Tint$ are given in the appendix of~\cite{CS2}.} Using the orthonormal basis of Killing spinors to split the $\SU(8)$ spinor indices as $\alpha = (a,i)$, it was also shown that the projection of the intrinsic torsion onto this bundle $V$ is constrained by a specific component of the algebra of Killing spinors $(\LgenS_{V_{ij}} \hat\epsilon_{k})^a \in V$, i.e. 
\begin{equation*}
(\LgenS_{V_{ij}} \hat\epsilon_{k})^a = 0 \,\Rightarrow \,\exists\, D:\, D\hat{\epsilon}_i=0,\, T(D)|_V = 0.
\end{equation*} 
But equation~\eqref{eq:KSA-coef} states that indeed $(\LgenS_{V_{ij}}  \hat\epsilon_k)^a = 0$, and the Killing spinor equations~\eqref{eq:kse-su8} imply that all the $S+J$ components also vanish~\cite{CS2}, except for a singlet~\cite{CS1} which is set to the cosmological constant $\Lambda$. This is precisely what we needed to show.

We therefore conclude that there is an equivalence between spaces with constant generalised singlet intrinsic $\SU(8-\mathcal{N})$ torsion and our class of $\text{AdS}_4$ flux backgrounds which preserve $\mathcal{N}$ supersymmetries. 
In analogy with the nomenclature for weak $G_2$ manifolds which are $G_2$ manifolds up to a singlet torsion, we say that these AdS backgrounds with generic fluxes are spaces with \textit{weak generalised special holonomy}.
It should also be clear that the same result will hold in other dimensions for the relevant $G_{\mathcal{N}}$-structure in $\Edd\times\bbR^+$ generalised geometry. 
The $\cN=2$ case of this result has already been studied extensively 
in~\cite{Ashmore:2016qvs}, where some applications to AdS/CFT are given (see also~\cite{Grana:2016dyl,Ashmore:2016oug}). 
The $\cN=4$ case has also appeared in~\cite{Malek1,Malek2} where it was used to demonstrate the existence of universal consistent truncations to pure $\cN=4$ gauged supergravity in the external space.
As mentioned before, these works 
label this as the general situation as they 
overlook the subtlety involving $S_{ij}$ that appears for example in~\eqref{eq:ks-gckv}, though this is not an issue for $d=4,6$ as there is no analogue of the functions $S_{ij}$ in those cases.
It is known that there exist compact $\cN=3$ solutions, for example the tri-Sasakian solutions discussed in~\cite{Acharya:1998db,Cassani:2011fu}, and the structure we describe here will also include these cases.

Note that the singlets in the torsion which become identified with the cosmological constant are such that they break the Minkowski R-symmetry to the AdS one, see table~\ref{table3}. For the $d=7$ case, it was noted in~\cite{CS1} that 
$\Lambda$ must be in the $\rep{36}$ of $\SU(8)$, since this part of the torsion is constrained by~\eqref{eq:kse-su8-dil}, but not by~\eqref{eq:kse-su8-grav}. This component of the torsion is a symmetric 2-tensor $\Sigma_{\alpha\beta}=\Sigma_{(\alpha\beta)}$. If we decompose under $\SU(8-\mathcal{N})\times U(\mathcal{N})\subset \SU(8)$, with indices $(a,i)=\alpha$ respectively, we have that $\Sigma_{\alpha\beta}=(\Sigma_{ab},\Sigma_{ai},\Sigma_{ij})$. We thus find that $\Sigma_{ij}\propto \Lambda \delta_{ij}$ are the torsion singlets of the reduced structure for $G_{\mathcal{N}}=\SU(8-\mathcal{N})$, and this is what is fixed by the AdS Killing spinor equations to be proportional to the constant $\Lambda$. The AdS R-symmetry $\SO(\cN)$ is the subgroup of $U(\mathcal{N})$ which stabilises this symmetric 2-tensor, thus the cosmological constant becomes a singlet of $\SU(8-\cN)\times\SO(\cN)$. 
In this case, there is also the complex conjugate singlet in the $\rep{\bar{36}}$ which appears in the conjugate equations and ensures that the torsion is real overall.

\begin{table}[htb]
\centering
\begin{tabular}{lllll}
   $d$ & $\dHd$ &  $G_{\mathcal{N}}$ & $\text{Mink}_D$ R-sym & $\text{AdS}_D$ R-sym $\AdSR$ \\
   \hline
   7 & $\SU(8)$ &  $\SU(8-\mathcal{N})$ & $\left\{ \begin{matrix} U(\mathcal{N}), \; \mathcal{N}\neq 8 \\ \SU(8), \; \mathcal{N}=8 \end{matrix} \right.$ & $\SO(\mathcal{N})$ \\[8pt]
   6 & $ \Symp(8)$ & $\Symp(8-2\mathcal{N})$ & $\Symp(2\mathcal{N})$ & $
   \left\{ \begin{matrix} U(\mathcal{N}), \; \mathcal{N}\neq 4 \\ \SU(4), \; \mathcal{N}=4 \end{matrix} \right. $ \\[18pt]
   5 & $
   \Symp(4)^2 $ &
      $\!\!\!  \begin{matrix}\Symp(4-2\mathcal{N}_+)\\ \;\; \times\!\! \Symp(4-2\mathcal{N}_-)
   \end{matrix}$ &  $\!\!\!  \begin{matrix}\Symp(2\mathcal{N}_+) \\ \;\; \times\!\! \Symp(2\mathcal{N}_-) \end{matrix}$  &  $\left\{ \begin{matrix} \Symp(2\mathcal{N}), & \mathcal{N_+}=\mathcal{N}_- \\  \text{---}, &\mathcal{N_+}\neq\mathcal{N}_- \end{matrix} \right. $ \\[18pt]
   4 & $\Symp(4)$ & $\Symp(4-2\mathcal{N}) $ & $\Symp(2\mathcal{N}) $ & $\Symp(2\mathcal{N}) $
\end{tabular}
\caption{AdS backgrounds with $\mathcal{N}$ supersymmetry correspond to generalised $G_{\mathcal{N}}$-structures with constant singlet torsion. The non-zero intrinsic torsion of the $G_{\mathcal{N}}$-structure breaks the Minkowski R-symmetry group, which is simply the commutant of $G_{\cN}$ inside $\dHd$, to the AdS symmetry $\AdSR$. Note that, unlike the Minkowski case, there are no chiral supersymmetric AdS$_6$ backgrounds, or equivalently, there are no singlets in the torsion of a $\Symp(4-2\mathcal{N}_+)\!\!\times\!\!\Symp(4-2\mathcal{N}_+)$ structure if $\mathcal{N_+}\neq\mathcal{N}_-$.}
\label{table3}
\end{table}

Similarly in dimensions $d<7$, one finds that there is a singlet of $G_{\cN}\times \AdSR$ in the decomposition of the torsion which is constrained by the dilatino equation~\eqref{eq:kse-su8-dil} but not the gravitino variation~\eqref{eq:kse-su8-grav}. Unlike the $d=7$ case, for $d<7$ the singlet is uniquely specified by these requirements. In the $\cN=1,2$ and $4$ cases, these singlets were also discussed in~\cite{CS1},~\cite{Ashmore:2016qvs} and~\cite{Malek1,Malek2} respectively.

Explicitly, for $d=6$, let $\alpha,\beta = 1,\dots, 8$ be $\Symp(8)$ indices. These split under the $\Symp(8-2\mathcal{N})\times\SU(\mathcal{N})$ subgroup as $(X_\alpha) \ra (X_i,X^i,X_a)$ where $i=1,\dots,\cN$ is an $\SU(\cN)$ index and $a=1,\dots,8-2\cN$ is an $\Symp(8-2\cN)$ index. Note that the indices $i,j$ label the complex Killing spinors $\epsilon_i = \epsilon_i^1$, which are the first ``halves" of the internal symplectic Majorana spinors (as explained in section~\ref{sec:KSE}). One can think of the other ``halves" as being labelled by the conjugate indices (i.e. as $\epsilon^i = \epsilon_i^2$).
There is a component of the torsion $\Sigma_{(\alpha\beta)}$ in the $\rep{36}$ of $\Symp(8)$, and this contains a unique singlet $\Sigma_{i}{}^j \sim \Lambda \delta_i{}^j$ under the $\Symp(8-2\mathcal{N})\times U(\cN)$ (or $\SU(4)$ for $\cN=4$) subgroup, and it is this which becomes the cosmological constant. 

For $d=5$, one has two copies of the $\repp{4}{4}$ representation of $\Symp(4)\times\Symp(4)$ in the torsion. Letting $\alpha,\beta = 1,\dots,4$ and $\bar\alpha,\bar\beta = 1,\dots,4$ be indices for the two $\Symp(4)$ groups, these correspond to tensors, 
\begin{equation}
	\Sigma_{\alpha\bar\alpha}{}_{(\beta \gamma)} 
		= \Sigma_{\bar\alpha(\beta} C_{\gamma)\alpha}
	\hs{15pt} \text{and} \hs{15pt}
	\Sigma_{\alpha\bar\alpha}{}_{(\bar\beta\bar\gamma)} 
		= \tilde\Sigma_{\alpha(\bar\beta} C_{\bar\gamma)\bar\alpha} ,
\end{equation}
when one writes them with the index structure of the generalised connection, where $C_{\alpha\beta}$ and $C_{\bar\alpha\bar\beta}$ are the natural antisymmetric $\Symp(4)$ invariants. A short calculation then reveals that the singlet for the cosmological constant can be written as
\begin{equation}
	\Sigma_{ij}
	= -\tilde\Sigma_{\alpha\bar\alpha} 
	\sim  \Lambda \epsilon_{ij},
\end{equation}
where we have decomposed the barred and unbarred $\Symp(4)\times\Symp(4)$ indices as $\alpha \ra (a,i)$ and $\bar\alpha \ra (\bar{a},i)$. Here, $a=1,2$ and $\bar{a}=1,2$ are barred and unbarred $\Symp(2)\times\Symp(2)$ indices and $i=1,2$ are $\SU(2)$ indices, with the same $\SU(2)$ group appearing in both $\Symp(4)$ decompositions. $\epsilon_{ij}$ is simply the usual anti-symmetric invariant of $\SU(2)$. 

For $d=4$, the relevant singlet is simply the $\Symp(4)$ singlet in the torsion, which does not break the Minkowksi R-symmetry group as it is already invariant.

Finally, we have our conclusion:
\begin{quote}Given a $d$-dimensional space with arbitrary fluxes and a set of $\mathcal{N}$ orthonormal spinors $\hat\epsilon_i$, we have that it is a valid $\mathcal{N}$-supersymmetric $\text{AdS}_{D}$ flux background with cosmological constant $\Lambda$ \textit{if and only if }there exists a generalised torsion-free connection $D$ such that $(D-\Lambda\cdot)\epsilon_i = 0$ for all $\epsilon_i$ and the torsion $\Lambda$ is a particular constant singlet of 
$G_{\cN}\times\AdSR$, as detailed above.
\end{quote}
As reasoned at the start of this section, this statement is also valid, over $(d-1)$-dimensional space, for the description of type II theories in $\Edd\times\bbR^+$ generalised geometry.


\section{Killing superalgebra}
\label{sec:ksa}

The formalism of generalised geometry, and in particular the Kosmann-Dorfman bracket~\eqref{eq:gen-KD-su8}, also allows us to compute the algebra generated by the Killing spinors of AdS flux backgrounds and its coefficients. 
In fact, these also fix the higher-dimensional Killing superalgebra, which is the Lie superalgebra of Killing spinors and Killing vectors of the eleven-dimensional background introduced in~\cite{FigueroaO'Farrill:2004mx}. 
As for the Minkowski solutions, the algebra turns out to have some remarkable properties which, even though not necessarily unexpected, would be difficult to demonstrate using other methods.

We will perform explicit calculations only in the AdS$_4$ case, and make some more general comments on other dimensions at the end.
As in section~\ref{sec:gen-hol}, all of the calculations and results found at the level of $\dHd$ representations will be equally true for the $\Edd\times\bbR^+$ generalised geometry descriptions of type II theories, though our presentation here is mainly focused around backgrounds of eleven-dimensional supergravity.

\subsection{The AdS$_4$ case}
\label{sec:AdS4-KSA}

Repeating the calculation of~\cite{CS2}, we may compute the coefficients $X_{ijk}{}^l$ of the Kosmann-Dorfman bracket of the Killing spinors in~\eqref{eq:KSA-coef} to find 
\begin{equation}
	L_{V_{ij}}\epsilon_k = \tfrac{3}{32} D_{V_{[ij}} \epsilon_{k]} 
		+ \tfrac1{16} \Lambda \delta_{k[i} \epsilon_{j]}
		= \tfrac1{16} \Lambda \delta_{k[i} \epsilon_{j]} .
\end{equation}
This is the adjoint action of $\SO(\cN)$ on the fundamental representation, which corresponds to
\begin{equation}
	X_{ijk}{}^l = \tfrac1{16} \Lambda \delta_{k[i} \delta_{j]}{}^l .
\end{equation}
We see that the Jacobi identity for the algebra of the Killing spinors is clearly satisfied, since it corresponds to $X_{[ijk]}{}^l=0$. Note also that $\delta_{ij}$ appears in these formulae, and as such they are only invariant under $\SO(\mathcal{N})$ rotations of the Killing spinors. This is because the cosmological constant is breaking the R symmetry to $\SO(\mathcal{N})$, 
of which the Killing spinors form a real representation.

Similarly, for $W^{ij}$ we find
\begin{equation}
	L_{W^{ij}}\epsilon_k = \tfrac1{16} \Lambda \delta_{k[i} \epsilon_{j]} .
\end{equation}
Again, this is only $\SO(\mathcal{N})$ covariant on the $i,j$ indices labelling the basis of Killing spinors. From now on, we will not carefully distinguish between the upstairs and downstairs $i,j$ indices for convenience, and it is understood that they are raised and lowered with $\delta_{ij}$.

In fact, it is better to write all of this in terms of \textit{real} generalised Killing vectors $U_{ij}$ and $\tU_{ij}$ such that
\begin{equation}
\label{eq:V-U-tU}
	V_{ij} = U_{ij} + \ii \tU_{ij} .
\end{equation}
One finds
\begin{equation}
\label{eq:int-KSA-1}
\begin{aligned}
	L_{U_{ij}}\epsilon_k &= \tfrac1{16} \Lambda \delta_{k[i} \epsilon_{j]} ,
	& \hs{40pt} &&
	L_{\tU_{ij}}\epsilon_k &= 0 , \\
	L_{U_{ij}}\bar\epsilon_k &= \tfrac1{16} \Lambda \delta_{k[i} \bar\epsilon_{j]},
	& \hs{40pt} &&
	L_{\tU_{ij}}\bar\epsilon_k &= 0 .
\end{aligned}
\end{equation}
As these are given by the adjoint action of the $\SO(\mathcal{N})$ algebra, we must have then
\begin{equation}
\label{eq:AdS4-isom}
\begin{aligned}
	L_{U_{ii'}} U_{jj'} &= \tfrac18 \Lambda \delta_{ij} U_{i'j'} ,
	& \hs{40pt} &&
	L_{\tU_{ii'}} U_{jj'} &= 0 ,\\
	L_{U_{ii'}} \tU_{jj'} &= \tfrac18 \Lambda \delta_{ij} \tU_{i'j'} ,
	& \hs{40pt} &&
	L_{\tU_{ij}} \tU_{jj'} &= 0 ,
\end{aligned}
\end{equation}
where we understand that $ii' = [ii']$ are antisymmetrised. The fact that the algebra of these vectors closes with constant structure constants suggests that they will give rise to the gauge algebra of a universal consistent truncation of the type studied in~\cite{Gauntlett:2007ma,Gauntlett:2006ai,Gauntlett:2010vu,Passias:2015gya}. For $\mathcal{N}=8$, this is precisely the algebra of the generalised parallelisation of the $S^7$ solution~\cite{Lee:2014mla} and for the half-maximal cases, this has recently been shown in~\cite{Malek1,Malek2}.

From~\eqref{eq:AdS4-isom}, we see that the generalised Killing vectors $U_{ij}$ satisfy the $\SO(\mathcal{N})$ algebra. We can now employ a simple argument (c.f. the no-go theorem of~\cite{LSW2}), to show that this implies the existence of ordinary Killing vectors (i.e. Riemannian isometries) with an $\SO(\mathcal{N})$ algebra. 
The generalised tangent space $E\simeq TM \oplus \Lambda^2T^*M \oplus \Lambda^5T^*M \oplus\dots$ has a graded structure, i.e. $T$ has degree zero, $\Lambda^2T^*$ has degree one, etc. Writing a generalised vector \[V= v + \omega+ \sigma +\dots \in TM \oplus \Lambda^2T^*M  \oplus \Lambda^5T^*M\oplus \dots\] then the Dorfman derivative takes the form~\cite{PW,CSW2} \[L_V V' = \mathcal{L}_vv' + \left( \mathcal{L}_v\omega ' - i_{v'}\dd \omega \right)+ \left( \mathcal{L}_v \sigma' - i_{v'} \dd\sigma
          - \omega'\wedge\dd\omega \right) + \dots\] and it is clear that it is compatible with the grading. Note in particular that the vector component is just the Lie bracket of the ordinary vectors. 

Now let $\mathcal{G}$ be the (constant) real linear span of the generalised Killing vectors $U_{ij}$ and $\mathcal{I}\subset \mathcal{G}$ be the subspace consisting of those generalised vectors with zero vector component.  By definition, $\mathcal{I}$ is in the span of the sub-bundles of strictly positive degree  
and is thus a solvable (nilpotent, in fact) ideal of the algebra generated by the $U_{ij}$ under the Dorfman bracket. However, the $\SO(\mathcal{N})$ algebra is semi-simple, and thus $\mathcal{I}$ must be zero. Therefore, all of the generalised vectors in the span of the $U_{ij}$ have a non-trivial vector part and, again from the graded structure of the Dorfman bracket, these must also form the $\SO(\mathcal{N})$ algebra in the Lie bracket. Thus we have an $\SO(\mathcal{N})$ algebra of infinitesimal isometries as claimed.\footnote{Though we referred to the $\GL(d,\bbR)$ decomposition of the generalised tangent space here, the same graded structures are also present in the type II decompositions and the same arguments can be used.}

\subsubsection{Eleven-dimensional interpretation}

Similarly to~\cite{CS2}, we now construct a collection of eleven-dimensional spinors using a basis $\eta_A$ of Majorana Killing spinors on AdS$_4$ and our basis $\epsilon_i$ of internal Killing spinors. It is natural to construct the real eleven-dimensional spinors\footnote{See~\cite{CSW3} for a full account of our spinor conventions.}
\begin{equation}
\label{eq:11d-spinor-basis}
	Q_{Ai} = \eta_{A+} \otimes \epsilon_i + \eta_{A-} \otimes \epsilon_i^* ,
\end{equation}
which form a real representation of $\SO(3,2)\times\SO(\mathcal{N})$. Here, $A = 1,\dots,4$ is the fundamental representation of $\Symp(4,\bbR) \simeq \SO(3,2)$ while $i=1,\dots,\mathcal{N}$ labels the fundamental representation of $\SO(\mathcal{N})$ as before.
The AdS$_4$ Killing spinors satisfy
\begin{equation}
	\LC_\mu \eta_A = \tfrac12 \Lambda \gamma_\mu \eta_A ,
\end{equation}
giving rise to 
\begin{equation}
	k^\mu_{(AB)} = \bar\eta_A \gamma^\mu \eta_B, 
\end{equation}
which are the $\SO(3,2)$ Killing vectors on AdS$_4$. 
Their inner products also provide the $\Symp(4,\bbR)$ invariant, 
\begin{equation}
	\bar\eta_A \eta_B = C_{AB} = \text{constant.} 
\end{equation}

For the internal Killing spinors, from~\eqref{eq:ks-innerprod}, the inner products are (recall that we assume $S_{ij}=0$ here)
\begin{equation}
	\epsilon^\dagger_i \epsilon_j = \ee^\Delta \delta_{ij} ,
\end{equation}
and we define internal vectors $u_{ij}$ and $\tu_{ij}$ by setting 
\begin{equation}
	\epsilon^T_i \gamma^m \epsilon_j = u^m_{ij} + \ii \tu^m_{ij} .
\end{equation}
Up to a rescaling, these are the (coordinate frame) vector components of the generalised vectors $U_{ij}$ and $\tU_{ij}$ from~\eqref{eq:V-U-tU}.

In terms of these objects, the vector bilinear of the Killing spinors becomes
\begin{equation}
\label{eq:QQ-AdS4}
	\{ {Q}_{Ai},  Q_{Bj} \} = k_{AB} \delta_{ij} + C_{AB} u_{ij} .
\end{equation}
This is the $\{ \text{Odd}, \text{Odd} \} \!\!\ra\!\! \text{Even}$ bracket for the $OSp(4|\cN)$ superalgebra. 
The $k_{AB}$ and $u_{ij}$ span the  even subalgebra $\SO(3,2)\times\SO(\mathcal{N})$ by~\eqref{eq:AdS4-isom}, and using~\eqref{eq:int-KSA-1} we see that these act on the odd subspace spanned by the ${Q}_{Ai}$ as the natural adjoint action of $\SO(3,2)\times\SO(\mathcal{N})$ on the $\repp{4}{\mathcal{N}}$ representation. Thus, we have fixed the form of the eleven-dimensional Killing superalgebra. It is isomorphic to $OSp(4|\cN)$, the usual $\mathcal{N}$-extended AdS$_4$ superalgebra, as one would expect.

We stress again that the condition that $S_{ij}=0$ in~\eqref{eq:ks-innerprod} is important, as it is necessary to obtain the natural relation~\eqref{eq:QQ-AdS4} for the Killing superalgebra~\cite{CS4}.


\subsection{Brief comments on other dimensions}

The exact same construction of generalised Killing vectors (as in~\eqref{eq:gen-bilinears}) and the other parts of the internal Killing superalgebra (as in~\eqref{eq:int-KSA-1} and~\eqref{eq:AdS4-isom}) can be employed to find the corresponding results for AdS$_D$ backgrounds, for $D>4$. One can further show that there are ordinary isometries of the background satisfying the AdS R-symmetry algebra using the same arguments as in~\ref{sec:AdS4-KSA}. 
These isometries are often assumed by authors motivated by AdS/CFT. For example, the analysis of AdS$_6$ solutions in~\cite{DHoker} begins with an ansatz containing an $S^2$ factor on which the $\SU(2)$ isometry acts, though this was already shown to be a general feature of such solutions by direct calculation~\cite{GCG-AdS6}. 
In each case, one simply builds eleven-dimensional Killing spinors from tensor products of the basis of internal Killing spinors and a suitable basis of external AdS Killing spinors, as in~\eqref{eq:11d-spinor-basis}. These can then be seen to satisfy the relevant AdS superalgebras~\cite{Nahm:1977tg}, which are shown in table~\ref{table4}. 
\begin{table}[htb]
\centering
\begin{tabular}{l|cccc}
   $D$ & 4 & 5 & 6 & 7 \\
   \hline
   $\text{AdS}_D$ superalgebra 
   & $OSp(4|\mathcal{N})$
   & $SU(2,2|\mathcal{N})$
   &  $F(4)$
   & $OSp(6,2|\mathcal{N}) $ \\[5pt]
   Even subalgebra 
   	& $\begin{matrix} \SO(3,2) \\  
		\times \SO(\mathcal{N}) \end{matrix}$
	& $\left\{ \begin{matrix} \SO(4,2)\!\!\times\!\! U(\mathcal{N}), & \mathcal{N}<4 \\  
		\SO(4,2)\!\!\times\!\!\SU(4), & \mathcal{N}=4 \end{matrix} \right\} $
	& $\begin{matrix} \SO(5,2) \\  
		\times \SU(2) \end{matrix}$
	& $\begin{matrix} \SO(6,2) \\  
		\times \Symp(\cN) \end{matrix}$
\end{tabular}
\caption{The AdS superalgebras and their even subalgebras}
\label{table4}
\end{table}

Some of the subtleties of AdS superalgebras can be explained easily from the generalised geometry viewpoint. 
As an example, consider the the AdS$_5$ case, where the R-symmetry is $U(\mathcal{N})$ for $\mathcal{N}<4$, but $\SU(4)$ for $\mathcal{N}=4$. 
This can be understood by examining the decomposition of the generalised tangent space in $E_{6(6)}\times\bbR^+$ generalised geometry under the relevant subgroups (see table~\ref{table3}). The generalised vectors transform in the $\rep{27_{+1}}$ representation, which remains irreducible under the $\Symp(8)$ generalised metric subgroup.
We work in $\Symp(8)$ indices $\alpha,\beta = 1,\dots, 8$ and decompose them under $\Symp(8-2\cN)\times\SU(\cN)$ as in section~\ref{sec:weak-gen-hol}.
The generalised tangent space decomposes as \\
\begin{equation}
\label{eq:AdS5-decomp-a}
	V_{[\alpha\beta]} \ra (V_{[ab]}, V_{ai}, V_a{}^i, V_i{}^j, V_{[ij]}, V^{[ij]}) ,
\end{equation}
such that these components satisfy the symplectic traceless condition on the $\rep{27}$ representation of $\Symp(8)$
\begin{equation}
\label{eq:AdS5-decomp-b}
	C^{ab} V_{ab} + V_i{}^i = 0 .
\end{equation}
The singlets of the $\Symp(8-2\cN)$ structure are thus $(V_i{}^j, V_{ij}, V^{ij})$ and it is these parts which can be constructed as the bilinears of the Killing spinors $(\epsilon_i,\epsilon^i)$. Similarly to the result~\eqref{eq:AdS4-isom} above, we find that the $V_i{}^j$ give isometries of the internal space, which correspond to the AdS R-symmetry. These obey the $U(\cN)$ algebra for $\cN<4$. However, for $\cN=4$, the tracelessness condition~\eqref{eq:AdS5-decomp-b} degenerates to $V_i{}^i = 0$, and one has only 15 generalised vectors obeying the $\SU(4)$ algebra rather than the full $U(4)$.


\begin{acknowledgments}

We thank Davide Cassani, Stefanos Katmadas, Ruben Minasian and Dan Waldram for helpful discussions. 
The work of C.~S-C has been supported by the European Research Council ({\sc Erc}) under the EU Seventh Framework Programme (FP7/2007-2013) / {\sc Erc} Starting Grant (agreement n. 278234 - {\sc `NewDark'} project). 
A. C. has been supported by the Laboratoire d'Excellence CARMIN and the Agence
Nationale de la Recherche under the grant 12-BS05-003-01, and would like to thank the Institut des Hautes \'Etudes Scientifiques for hospitality during the completion of this work.
C.~S-C would like to thank the BIRS workshop ``String and M-theory geometries: Double Field Theory, Exceptional Field Theory and their Applications" (17w5018) and the COST Action MP 1405 meeting for hospitality during the completion of this work.

\end{acknowledgments}



\end{document}